\documentclass[final,times,5p,twocolumn]{elsarticle}

\usepackage{graphics}
\usepackage{subfigure,graphicx}
\usepackage{amssymb}
\usepackage{amsthm}
\usepackage{color}
\usepackage{hyperref}
\usepackage{booktabs}
\usepackage{amsmath}
\usepackage{mathrsfs}
\usepackage{mathtools, cuted}
\usepackage{lipsum, color}
\usepackage{multirow}

\newcommand{\gsim}{\gtrsim}
\newcommand{\lsim}{\lesssim}

\newcommand{\lf}{\left(}
\newcommand{\ri}{\right)}

\newcommand{\nn}{\nonumber}

\newcommand{\sqs}{\sqrt{s}}

\renewcommand{\lg}{\mathscr{L}}

\newcommand{\mco}{\mathcal{O}}

\newcommand{\br}{\mathcal{B}}

\newcommand{\sm}{{\rm SM}}
\newcommand{\tot}{{\rm tot}}

\newcommand{\fb}{{\;{\rm fb}}}

\newcommand{\iab}{{\;{\rm ab}^{-1}}}

\newcommand{\gev}{{\;{\rm GeV}}}
\newcommand{\tev}{{\;{\rm TeV}}}

\newcommand{\beq}{\begin{equation}}
\newcommand{\eeq}{\end{equation}}
\newcommand{\bea}{\begin{eqnarray}}
\newcommand{\eea}{\end{eqnarray}}
\newcommand{\barr}{\begin{array}}
\newcommand{\earr}{\end{array}}
\newcommand{\bc}{\begin{center}}
\newcommand{\ec}{\end{center}}
\newcommand{\bit}{\begin{itemize}}
\newcommand{\eit}{\end{itemize}}
\newcommand{\ben}{\begin{enumerate}}
\newcommand{\een}{\end{enumerate}}

\newcommand{\Dt}{\Delta}

\newcommand{\sg}{\sigma}

\newcommand{\kp}{\kappa}

\newcommand{\lm}{\lambda}

\newcommand{\bb}      {{b \bar{b}}}

\newcommand{\kpv} {\kappa_V}

\newcommand{\kptv} {\kappa_{2V}}
\newcommand{\kplm} {\kappa_\lambda}
\definecolor{mint}{rgb}{0.24, 0.71, 0.54}

\journal{Physis Letters B}

\begin{document}

\begin{frontmatter}



\title{Studies of nonresonant Higgs pair production at electron-proton colliders} 

\author{Adil Jueid}
\ead{adil.hep@gmail.com}
\author{Jinheung Kim}
\ead{jinheung.kim1216@gmail.com}
\author{Soojin Lee}
\ead{soojinlee957@gmail.com}
\author{Jeonghyeon Song}
\ead{jhsong@konkuk.ac.kr}

\address{Department of Physics, Konkuk University, Seoul 05029, Republic of Korea}

\begin{abstract}
The measurement of the Higgs quartic coupling modifier between a Higgs
boson pair and a vector boson pair, $\kappa _{2V}$, is expected to be achieved
from vector-boson fusion (VBF) production of a Higgs boson pair. However,
this process involves another unmeasured parameter, the trilinear Higgs
self-coupling modifier $\kappa _{\lambda }$. A sensitivity analysis should
target both parameters. Since the LHC cannot avoid the gluon fusion pollution,
which becomes severe for non-SM $\kappa _{\lambda }$, an electron-proton
collider is more appropriate for the comprehensive measurement. In this
regard, we study the VBF production of a Higgs boson pair in the
$b\bar{b}b\bar{b}$ final state at the LHeC and FCC-he. Performing detailed
analysis using the simulated dataset, we devise the search strategy specialized
at the LHeC and FCC-he and give a prediction for the sensitivity to both
$\kappa _{2V}$ and $\kappa _{\lambda }$. We find that the two electron-proton
colliders have high potential: the LHeC has similar exclusion prospects
as the HL-LHC; the FCC-he is extremely efficient, excluding the parameter
space outside $\kappa _{2V} \in [0.8, 1.2]$ and
$\kappa _{\lambda }\in [1, 2.5]$ at 95\% C.L. for the total luminosity of
$10$ ab$^{-1}$ and 10\% uncertainty on the background yields.
\end{abstract}


\end{frontmatter}

\section{Introduction}
\label{sec:intro}
Albeit the absence of any signatures of the physics beyond the Standard Model (BSM), 
the journey to the final theory of the Universe will never stop.
One important task to achieve the goal is to measure every coupling among the SM particles precisely, 
especially to the Higgs boson $H$. 
The Higgs coupling modifiers associated with a single Higgs boson
have been observed to be SM-like at the LHC~\cite{Aad:2019mbh,CMS:2020gsy}. 
Their future projections at the high luminosity LHC (HL-LHC) 
expect the precisions at or below the percent level~\cite{Cepeda:2019klc}.  
However, coupling modifiers involving a pair of Higgs bosons remain unmeasured,
such as $\kplm$ for the trilinear Higgs self-coupling
and $\kptv$ for the quartic coupling between a Higgs boson pair and a vector boson pair.
The $\kplm$ shall be probed mainly from nonresonant Higgs boson pair ($HH$) production via gluon fusion: the triangle diagram mediated
by the Higgs boson in the $s$-channel gives access to $\kplm$. 
It is found that if $\kplm \neq 1$, non-trivial changes occur 
on both the shape and rate of the main kinematic distributions~\cite{Cheung:2020xij}. The current observed interval at the 95\% confidence level (C.L.) is $-5.0 < \kp_\lm < 12.0$ in the ATLAS analysis~\cite{Aad:2019uzh} and $-11.8 < \kp_\lm < 18.8$ in the CMS analysis~\cite{Sirunyan:2018ayu}. 
Several studies of the prospects for measuring $\kplm$ at the HL-LHC and future colliders have been performed~\cite{Adhikary:2017jtu, Chang:2018uwu, Homiller:2018dgu, Li:2019jba, Park:2020yps, Mangano:2020sao, Amacker:2020bmn, Abdughani:2020xfo},
which expect more stringent bounds.
 
The quartic coupling modifier $\kptv$ is much more challenging to measure at the LHC
since the most efficient process, nonresonant $HH$ production via vector boson fusion (VBF),
has very small cross-section of $\left. \sg_{\rm VBF}^\sm(pp \to HHjj) \right|_{\rm N^3LO}=1.73\fb$ 
at $\sqs=13\tev$~\cite{Dreyer:2018qbw} in addition to the huge SM backgrounds.
The ATLAS collaboration performed the first search
and excluded $\kptv<- 0.76$ and $\kptv >2.90$ at the 95\% C.L.
for $\kp_V=1$ and $\kplm=1$~\cite{Aad:2020kub}.
The assumption of $\kp_V =  1$ 
is well motivated by the Higgs precision measurements at the LHC, but
$\kplm=1$ is questionable.
The VBF production of $HH$,
which also depends on $\kplm$ via the $H$-mediated $s$-channel diagram,
is susceptible to anomalous Higgs self-coupling ($\kplm\neq 1$).
For example, the cross-section for $\kplm=5$ at the 14 TeV LHC
is about twenty times that for $\kplm=1$.
More serious is the pollution from the gluon fusion production of $HH$ associated with two jets, 
$gg \to HHjj$~\cite{Dolan:2015zja,Dolan:2013rja}.
This pollution also has the contribution from $\kplm$
and greatly increases for $\kplm \neq 1$.\footnote{The ATLAS collaboration
treated the gluon fusion pollution as a background 
because they assumed $\kplm=1$ and thus knew its rate~\cite{Aad:2020kub}.}
Considering huge QCD uncertainties in the gluon fusion pollution~\cite{Baglio:2012np},
we expect an inevitable limitation to the precision measurement of $\kptv$ at the LHC.

Targeting the measurements of $\kplm$ and $\kptv$ 
without the assumption about $\kplm$ 
and thus the ambiguity of the gluon fusion pollution, we turn to two electron-proton colliders, 
the Large Hadron electron Collider (LHeC)~\cite{AbelleiraFernandez:2012cc,Bruening:2013bga,Agostini:2020fmq} 
and the Future Circular Collider (FCC-he)~\cite{Abada:2019lih}.
The development of the energy recovery linac for the electron beam 
makes it possible 
to simultaneously operate the $pp$ and $e^- p$ collisions.
In particular, the LHeC has a bright outlook
as its working group recently announced the default configuration and staging
based on the cost estimation~\cite{Agostini:2020fmq}.
We find the following advantages of electron-proton colliders in probing rare BSM events:
\bit
\item The pileup, which degrades the quality of the data for physics analyses, is very small even at the high luminosity option ($\sim 10^{34}/{\rm cm}^2/{\rm s}$): 
one expects about $0.1$~(1) pileup collisions per event at the LHeC (FCC-he) 
while $\gsim 150$ at the LHC.
\item The QCD backgrounds and the higher-order corrections are suppressed,
providing a clean environment.
\item The charged-current (CC) and neutral-current (NC) processes
can be disentangled by tagging the outgoing neutrino (as large missing transverse energy) or electron. Independent measurements of $\kp_{2W}$ and $\kp_{2Z}$ are possible.
\item The asymmetric initial state allows us to distinguish the forward and backward directions, 
which can increase the signal significance. 
\item High polarization of the electron beam, $P_e$, is feasible,
as large as $\pm 80\%$~\cite{Agostini:2020fmq}.
The CC production cross-section increases
by the factor of $(1-P_e)$,
while the NC cross-section does not change much.
\eit

For the configurations of
\bea
\label{eq:configuration}
\hbox{LHeC:} \!\!&&\!\! E_e=50\gev, \quad E_p=7\tev, 
\\ \nn
\hbox{FCC-he:} \!\!&&\!\! E_e=60\gev, \quad E_p=50\tev, 
\eea
we shall analyze the sensitivity of the LHeC and FCC-he to $\kptv$ and $\kplm$
via the VBF production of $HH$ through the CC channel.\footnote{In the literature, 
the $e^- p$ collider phenomenologies of anomalous Higgs couplings associated with a Higgs boson pair 
were studied in the effective Lagrangian model,
via the CC~\cite{Kumar:2015kca} and NC process~\cite{Kuday:2017vsh}.}
Taking full advantage of the characteristics of the electron-proton collider, 
we shall propose a search strategy which we believe is optimal for measuring $\kptv$ and $\kplm$.
Finally, we will present the 95\% C.L. exclusion in the $(\kptv,\kplm)$ space,
based on the detector-level analysis of the signals and the relevant backgrounds. 
The remainder of this letter is organized as follows. In section \ref{sec:Formalism}, we discuss the formalism of Higgs boson pair production in $e^-p$ collisions within the $\kappa$-framework along with a discussion of the modeling of the signal and background processes. 
In section \ref{sec:results}, we discuss the analysis strategy and present our results. We conclude in section \ref{sec:Conclusions}.

\section{Formalism and modeling for the signal and backgrounds}
\label{sec:Formalism}
Based on the observed Higgs precision data via single Higgs production at the LHC,
we assume that all the couplings to a \emph{single} Higgs boson are the same as in the SM:
\begin{equation}
\label{eq:single:kp}
\kp_{Hij} =1,
\end{equation}
where $i$ and $j$ are the SM particles.
For renormalizable couplings to a Higgs boson pair,
we consider
\begin{equation}
\label{eq:Lagrangian}
\lg \supset \kappa_{2V} \frac{g^2}{4} W^{+\mu} W^-_{\mu} H^2 - \kappa_\lambda \frac{3 m_H^2}{v} H^3,
\end{equation}
where $v\simeq 246\gev$.
Note that $\kptv$ and $\kplm$ parameterize the BSM interactions
within the context of the non-linear effective field theory given by the electroweak chiral 
Lagrangian~\cite{Appelquist:1980vg,Longhitano:1980iz,Dobado:1989ax,Dobado:1990zh,Herrero:1993nc}\footnote{Note that concrete BSM scenarios can accommodate large values of $\kplm$ at the quantum level \cite{Hollik:2001px, Kanemura:2004mg, Nhung:2013lpa, Arhrib:2015hoa,  Braathen:2019pxr}. 
However, it is not straightforward to construct a BSM model that can have a large $\kptv$ without significantly affecting $\kpv$.}.

Aiming at the precision measurement of $\kptv$ and $\kplm$ together,
we focus on the pair production of Higgs bosons through the CC VBF interaction 
in the $\bb\bb$ final state, 
\begin{equation}
\label{eq:signal}
    p e^- \to H H + j_{\rm f} \nu_e \to \bb\bb+ j_{\rm f} \nu_e,
\end{equation}
where $j_{\rm f}$ is a forward jet. 
There are 
three kinds of Feynman diagrams for this process,
the contact one involving $HHW^+ W^-$ coupling,
the $s$-channel involving $HHH$ coupling,
and the $t,u$-channels with the square of $HW^+ W^-$ coupling.
The scattering amplitudes of $W^+ W^- \to HH$ help us to understand 
the characteristics of the signal.
As explicitly shown in Ref.~\cite{Arganda:2018ftn},
the longitudinally polarized 
$W^+_L$ and $ W^-_L$ 
make an overwhelmingly dominant contribution.
The corresponding amplitude, $\mathcal{M}_{LL}$, in the limit of
$\sqs \gg m_H$ satisfies
\bea
\label{eq:MLL}
\frac{1}{g^2} \mathcal{M}_{LL} \!\!\! &=&  \!\!\! \lf \kptv- \kpv^2 \ri   \frac{s}{4 m_W^2}
+ \kplm \frac{3 m_H^2}{4 m_W^2}
\\ \nn &&\!\!\! 
+ \kpv^2 \left[ 1- \frac{m_H^2}{2 m_W^2} - \frac{2}{\sin^2\theta^*}\right]
 -\frac{\kptv}{2}  
 + \mco \lf \frac{m_{W,H}^2}{s} \ri,
\eea
where we keep the notation of $\kpv$ to show its effects
and $\theta^*$ is the scattering angle in the center-of-mass frame of $W^+ W^-$.
Note that the effect of $\kplm$ dominates in the small $m_{HH}(=\sqs)$ region
while that of $\kptv$ does in the high $m_{HH}$ region.

\begin{figure}[!h]
    \centering
    \includegraphics[width=0.85\linewidth]{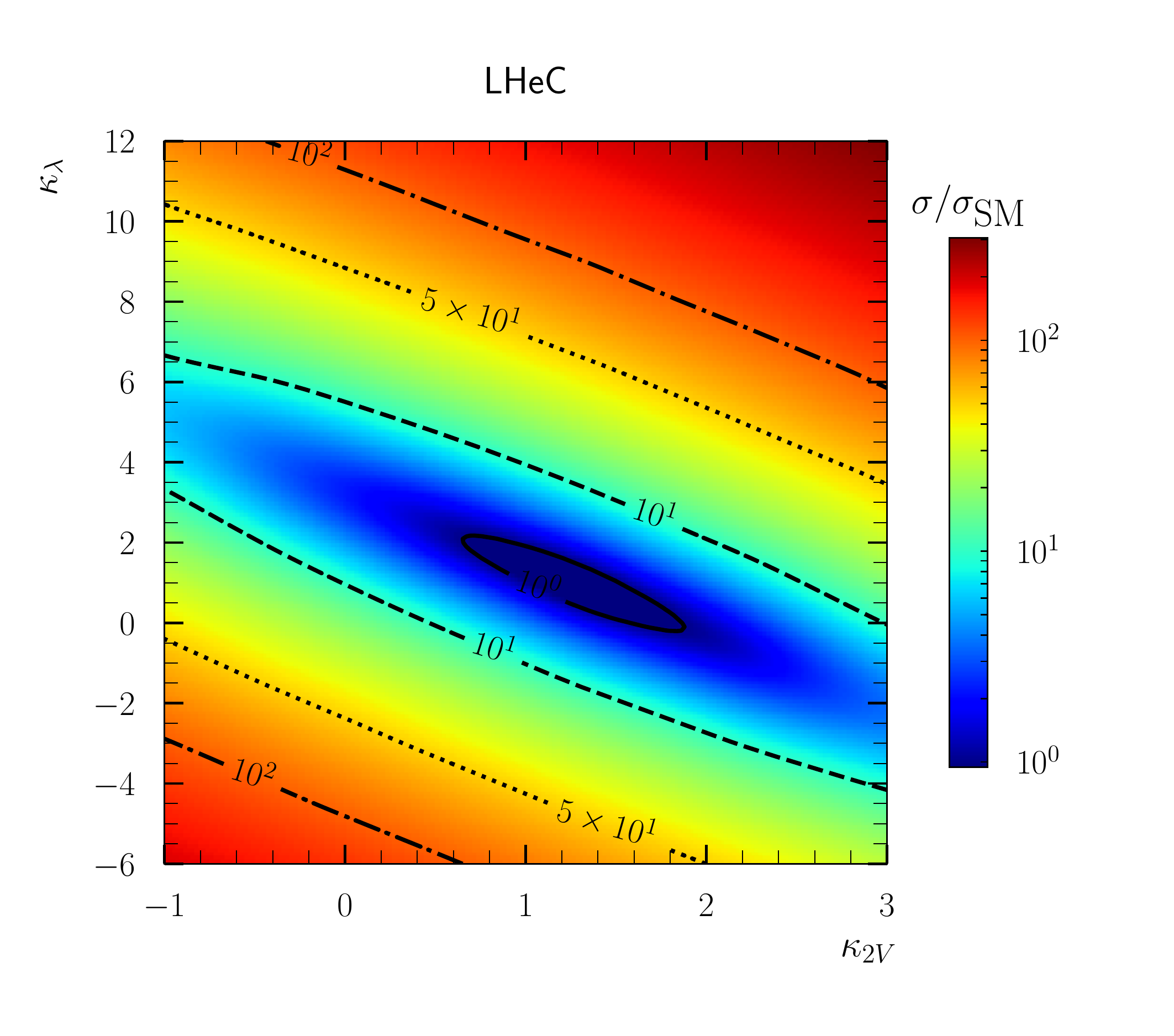}
    \includegraphics[width=0.85\linewidth]{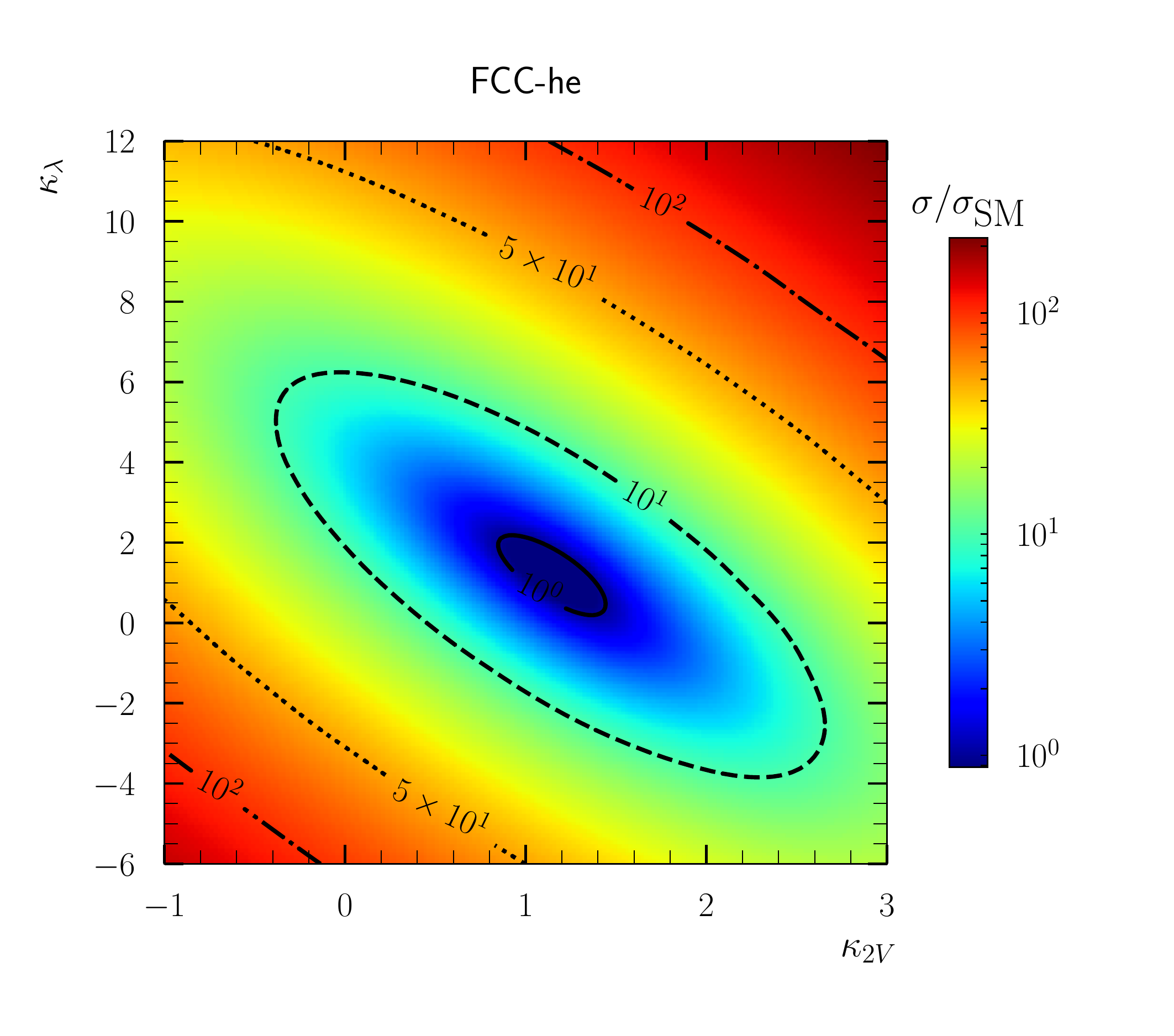}
    \caption{$\sigma/\sigma_{\rm SM}(p e^- \to HHj\nu_e)$ projected on 
    the plane of $(\kappa_{2V},\kappa_\lambda)$ at 
    the LHeC (upper panel) and FCC-he (lower panel). 
    The SM cross-section ($\kptv=\kplm=1$) with the unpolarized electron beam is 
    $\sigma_{\rm SM} = 5.97~{\rm ab}$ at the 
    LHeC and $\sigma_{\rm SM} = 233.77~{\rm ab}$ at the FCC-he.}
    \label{fig:signalXsec}
\end{figure}

First, at the parton level,
we calculate the total cross-sections of the signal 
by varying both $\kptv$ and $\kplm$.
The calculations have 
been performed at leading order (LO) using \texttt{MadGraph\_aMC@NLO}
with a modified \texttt{UFO}~\cite{Degrande:2011ua} model file 
for the Lagrangian in Eq.~(\ref{eq:Lagrangian}).
Based on the current experimental bounds,
we consider $-1 \leq \kappa_{2V} \leq 3$~\cite{Aad:2020kub}
and $ -6 \leq \kappa_\lambda \leq 12$~\cite{Aad:2019uzh,Sirunyan:2018ayu}.
The SM cross-section of the process $p e^- \to HHj_{\rm f}\nu_e$ is very small:
with the unpolarized electron beam, it is $\sigma_{\rm SM} = 5.97~{\rm ab}$ at the LHeC and 
$\sigma_{\rm SM} = 233.77~{\rm ab}$ at the FCC-he.
Despite tiny SM signals,
it is promising that the total cross-section rapidly increases 
when either $\kptv$ or $\kplm$ deviates from their SM values:
the two electron-proton colliders can exclude a large portion of the parameter space $(\kptv,\kplm)$.
To show this behavior,
we present $\sigma/\sigma_{\rm SM}$ in Fig.~\ref{fig:signalXsec}. 
It is clear to see that the deviation from $\kptv=1$ 
greatly increases the cross-section
because it invalidates the cancellation of the longitudinal polarization enhancement,
the first term of Eq.~(\ref{eq:MLL}).
The hypothesis of $\kplm\neq 1$ also increases the signal cross-section, 
though less than that of $\kptv\neq 1$. 
Quantitatively, we have a tenfold increase of $\sigma/\sigma_{\rm SM}$
if $ |\kptv-1|=|\kplm-1|=1$.
We also note that the same-sign $\kptv$ and $\kplm$ yield constructive interference,
explaining the negative slopes of the $\sigma/\sigma_{\rm SM}$ contours: see
the first two terms of Eq.~(\ref{eq:MLL}).
In detail, the LHeC and FCC-he show different shapes of the contours.
As shall be demonstrated, the LHeC is more sensitive to $\kplm$ than to $\kptv$.
The FCC-he has enough sensitivity to probe both.

\begin{table}[!t]
\setlength\tabcolsep{10pt}
    \centering
      {\renewcommand{\arraystretch}{1.5} 
    \begin{tabular}{r |c c}
    \toprule
    \multirow{2}{*}{Process} & \multicolumn{2}{c}{$\sg_{CC}$ [ab]} \\
        &    LHeC      &   FCC-he \\
    \toprule
    $b\bar{b}jj+j_{\rm f}\nu_e$ & $1.00\times 10^5\;\substack{+57.9\%\\  -33.9\%}$ & $7.18 \times 10^5\;\substack{+51.7\%\\ -31.6\%}$\\
    $ZZ+j_{\rm f}\nu_e$ &  $9.37\times 10^2\;\substack{+6.95\%\\ -5.45\%}$ & $2.24\times 10^4\;\substack{+3.65\%\\ -3.26\%}$\\
    $Zb\bar{b}+j_{\rm f}\nu_e$ & $5.38\times 10^2\;\substack{+27.5\%\\ -19.9\%}$ & $4.77 \times 10^3\;\substack{+22.6\%\\ -17.1\%}$ \\ 
    $ZH+j_{\rm f}\nu_e$ & $1.23 \times 10^2\;\substack{+7.29\%\\ -6.27\%}$ & $3.45 \times 10^3\;\substack{+3.99\%\\ -3.58\%}$ \\
    $b\bar{b}b\bar{b}+ j_{\rm f}\nu_e$     &    $1.82 \times 10^2\;\substack{+53.8\%\\ -32.4\%}$ &  $7.11 \times 10^2\;\substack{+50.9\%\\ -31.4\%}$ \\
    $Hb\bar{b}+j_{\rm f}\nu_e$ & $4.53 \times 10\;\substack{+28.4\%\\ -20.4\%}$ & $4.77 \times 10^2\;\substack{+23.7\%\\ -17.8\%}$\\
    $t\bar{t}+j_{\rm f}\nu_e$ & $2.00 \times 10\;\substack{+29.6\%\\ -21.2\%}$ & $7.49\times 10^2\;\substack{+22.9\%\\ -17.4\%}$ \\
    \bottomrule
    \end{tabular}
    }
    \caption{Parton level cross-sections in attobarn (ab) for the charged-current
background processes at the LHeC with $E_{e} = 50{\;{\mathrm{GeV}}}$ and $E_{p}
= 7{\;{\mathrm{TeV}}}$ and at the FCC-he with $E_{e} = 60{\;{\mathrm{GeV}}}$ and
$E_{p} = 50{\;{\mathrm{TeV}}}$. The electron beam is unpolarized. We have not included the decays of $Z$, $H$, and
the top quark. The uncertainties correspond to scale variations around the
nominal scale defined in Eq.~(\ref{eq:scales}). PDF uncertainties are at the
percent level and hence are not shown here.
\texttt{SysCalc}~\cite{Kalogeropoulos:2018cke} was used to compute these
uncertainties.
}
    \label{tab:bkgXsec}
\end{table}

For the $\bb\bb$ decay mode, the final state of the signal comprises at least four $b$-tagged
jets, one light untagged jet, and large missing transverse energy ($E_{T}^{\rm miss}$).
The main backgrounds\footnote{We do not present the NC backgrounds here
since they are to be highly suppressed by appropriate selection criteria.} are the QCD multi-jets, diboson, 
$t\bar{t}$, and single Higgs processes, 
all of which are associated with a forward jet $j_{\rm f}$
and an electron neutrino $\nu_e$.
In Table \ref{tab:bkgXsec}, 
we show the calculation of the LO cross-sections for the backgrounds at parton level
using \texttt{MadGraph\_aMC@NLO}~\cite{Alwall:2014hca} with \texttt{NNPDF31\_lo} parton distribution
function (PDF) set~\cite{Ball:2017nwa}.
Basic generator-level cuts were imposed on the parton-level 
objects like $p_T^j > 5~$GeV, $\Delta R_j > 0.4$, and $|\eta_j| < 10$.
The renormalization and factorization scales 
are set to be 
\bea
\label{eq:scales}
\mu_{R,0} = \mu_{F,0} \equiv \frac{1}{2} \sum_i \sqrt{p_{T,i}^2 + m_i^2}.
\eea
The total cross-section of all the CC backgrounds is about $100\fb$ ($751\fb$) at the LHeC (FCC-he).
The most dominant is the QCD production of  $b\bar{b}jj$,\footnote{In what follows,
we address each background process as the one without specifying $j_{\rm f} \nu_e$, for simplicity.}
where $j$ refers to a light quark (including a charm quark) or a gluon.
The second dominant backgrounds 
are from the production of a $Z$ boson associated with another $Z$ boson, the QCD $\bb$,
or a Higgs boson.
The QCD production of four $b$ quarks follows, and the production of a Higgs boson in association with $b\bar{b}$ is less critical.
Finally, the contribution
of a top quark pair production is smaller than the QCD $4b$ 
at the LHeC, but similar at the FCC-he. Important theoretical uncertainties arose from the scale variations,
as shown in Table \ref{tab:bkgXsec}. 
PDF uncertainties are of order $1$-$2\%$ for all the backgrounds.

We close this section by summarizing the Monte Carlo event generation procedure. Initially, events for the signal and backgrounds are generated at LO using \texttt{MadGraph\_aMC@NLO} version 2.6.7. Parton luminosities were modeled with the \texttt{NNPDF31\_lo} 
PDF set with $\alpha_s(m_Z^2) = 0.118$. Setting the direction of the proton beam as forward, we convolute the partonic cross-sections with the PDFs in the \texttt{LHAPDF6} library~\cite{Buckley:2014ana}. 
The decays of $H$, $Z$, and the top quark are modeled with \texttt{MadSpin}~\cite{Artoisenet:2012st}.
We confirmed that various kinematic distributions
from on-shell samples using \texttt{MadSpin} well agree with those from the off-shell samples.
For the parton-showering and hadronization,
we rely on \texttt{Pythia6}~\cite{Sjostrand:2006za}
since \texttt{Pythia8} does not support the LHE input 
in electron-proton collisions yet.
To correctly model hadronization of the events, we modified the default \texttt{Pythia6} setup. 
First, we switch off the lepton PDF by setting \texttt{MSTP(11)=0}. 
Second, we also switch off the QED initial state radiation for the electron beam by setting \texttt{MSTP(61)=0}. 
Finally, we switch off the negligible multiple-parton interactions, 
which saves a considerable amount of computing time. 
We use the default PDF at the \texttt{Pythia6} 
level,  \texttt{CTEQ6l}~\cite{Nadolsky:2008zw}. 
Fast detector simulation was performed
using \texttt{Delphes} version 3.4.2~\cite{deFavereau:2013fsa}. 
To match the particle efficiencies, momentum 
smearing, and isolation parameters with the default values in the Concept Design 
Report of the LHeC~\cite{Agostini:2020fmq},
we have performed minor modifications on the Delphes cards
in the GitHub repository \url{https://github.com/delphes/delphes/tree/master/cards}.
Jets are clustered using the anti-$k_T$ algorithm~\cite{Cacciari:2008gp} with a jet radius $R = 0.4$ in \texttt{FastJet} version 3.3.2~\cite{Cacciari:2011ma}. The $b$-tagging efficiency
is set to be $70\%$.
For the mistagging rates of the light and charm jets as a $b$ jet,
we adopted the default values in the above Delphes cards:
at the LHeC, $P_{j\to b} = 0.001$ and $P_{c\to b} = 0.05$;
at the FCC-he, 
$P_{j\to b} = 0.001$ and $P_{c\to b} = 0.04$ for $|\eta|<2.5$,
$P_{j\to b} = 0.00075$ and $P_{c\to b} = 0.03$ for $2.5 < |\eta| < 4$.

\section{Results and Discussion}
\label{sec:results}

\subsection{Event selection}

\begin{table*}[!t]
\setlength\tabcolsep{10pt}
\centering
{\renewcommand{\arraystretch}{1.1} 
\begin{tabular}{c |c c c c |r | c }
\toprule
 {Cut}    &  $b\bar{b}jj/\bb\bb$ & ~~$ZZ/HZ$~~ & $(Z/H)b\bar{b}$ & $t\bar{t}$ & total backgrounds  & Signal~($\kplm=\kptv=1$)  \\
\toprule
 & \multicolumn{6}{c}{LHeC with  $\mathcal{L}_\tot = 1~{\rm ab}^{-1}$ }\\
 \toprule
Initial & $100167.05$ &   $32.10$ & $107.41$ &   $17.65$ &  $100324.21~(100\%)$ & $1.98~(100\%)$\\
$4b$-tag & $4.36$   & $2.77$  & $2.26$ & $0.02$ &  $9.41~(0.0094\%)$ & $0.25~(12.34\%)$\\
Forward jet & $1.80$ & $2.15$ & $1.24$ & $0.01$ &  $5.20~(0.0052\%)$ & $0.17~(8.88\%)$ \\
Lepton veto & $1.80$ & $2.15$  & $1.24$ &  $0.01$ &   $5.20~(0.0052\%)$ & $0.17~(8.88\%)$ \\
$E_{T}^{\rm miss}>40\gev$ & $1.33$ &  $1.52$  & $0.95$ &   $0.01$ &   $3.81~(0.0038\%)$ & $0.074~(3.73\%)$ \\
Minimum $D_{HH}$ & $1.27$ &  $1.48$ &  $0.91$  &   $0.01$   & $3.66~(0.0037\%)$ & $0.064~(3.25\%)$ \\
$X_{HH} < 3.0$ & $0.15$ & $0.23$ & $0.17$ & $0.00$ & $0.55~(0.00043\%)$ & $0.04~(2.04\%)$ \\
\toprule
 & \multicolumn{6}{c}{FCC-he with  $\mathcal{L}_\tot = 10~{\rm ab}^{-1}$}\\
\toprule
Initial & $7180161$ & $8141.8$  & $9989.6$  & $6673.6$ & $7204970.0~(100\%)$ & $779.70~(100\%)$\\
$4b$-tag & $934.2$ &  $745.6$ & $274.7$ &  $24.2$  & $1978.7~(0.026\%)$ & $94.05~(12.06\%)$\\
Forward jet & $562.6$ &  $637.1$ & $185.6$ & $21.9$ &  $1407.2~(0.018\%)$ & $74.71~(9.58\%)$\\
Lepton veto & $562.5$ & $637.0$  & $185.6$ & $18.8$  & $1403.9~(0.018\%)$ & $74.71~(9.58\%)$ \\
$E_{T}^{\rm miss}>40\gev$ & $492.3$ & $497.7$ &  $153.0$ & $16.2$  & $1159.2~(0.015\%)$ & $42.11~(5.40\%)$ \\
Minimum $D_{HH}$ & $412.9$  & $458.8$  & $129.9$ &  $13.9$ &   $1015.5~(0.014\%)$ & $29.71~(3.81\%)$ \\
$X_{HH} < 2.0$ & $29.8$ &   $32.6$ & $10.2$ &  $1.5$  & $74.1~(0.00098\%)$ & $10.99~(1.41\%)$ \\
\bottomrule
\end{tabular}
}
\caption{Cut-flow chart of the number of events of the signal and backgrounds at the LHeC and 
the FCC-he with the unpolarized electron beam. 
The background processes are denoted as omitting $j_{\rm f} \nu_e$ for simplicity. 
The numbers inside the parentheses show the acceptance
times efficiency after the selection step $i$
with respect to the initial number of events $n_0$, \emph{i.e.} $\epsilon_i  = n_i/n_0$.
}
\label{tab:Efficiency}
\end{table*}

\begin{figure}[!h]
    \centering
    \includegraphics[width=0.9\linewidth]{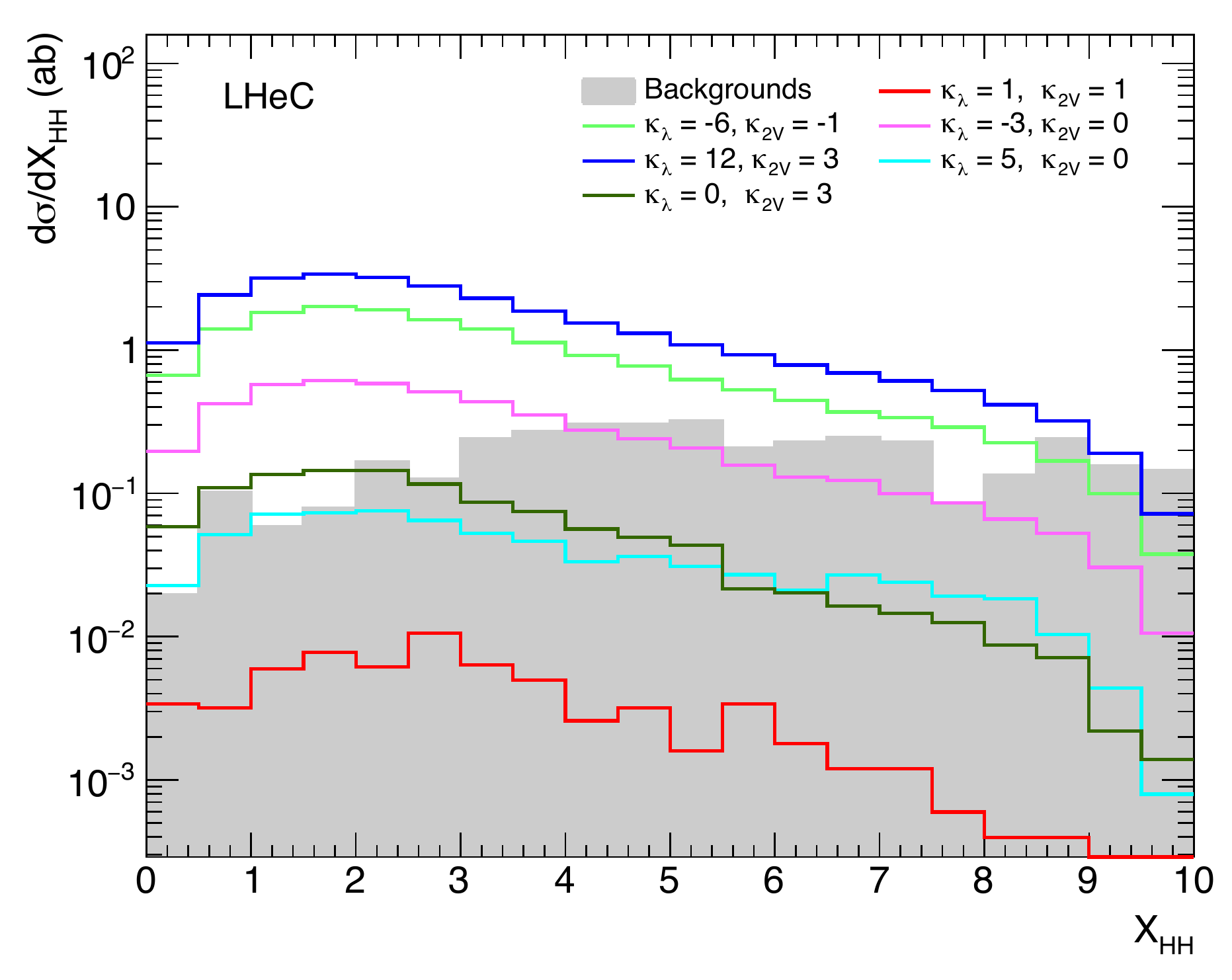}
    \hfill
    \includegraphics[width=0.9\linewidth]{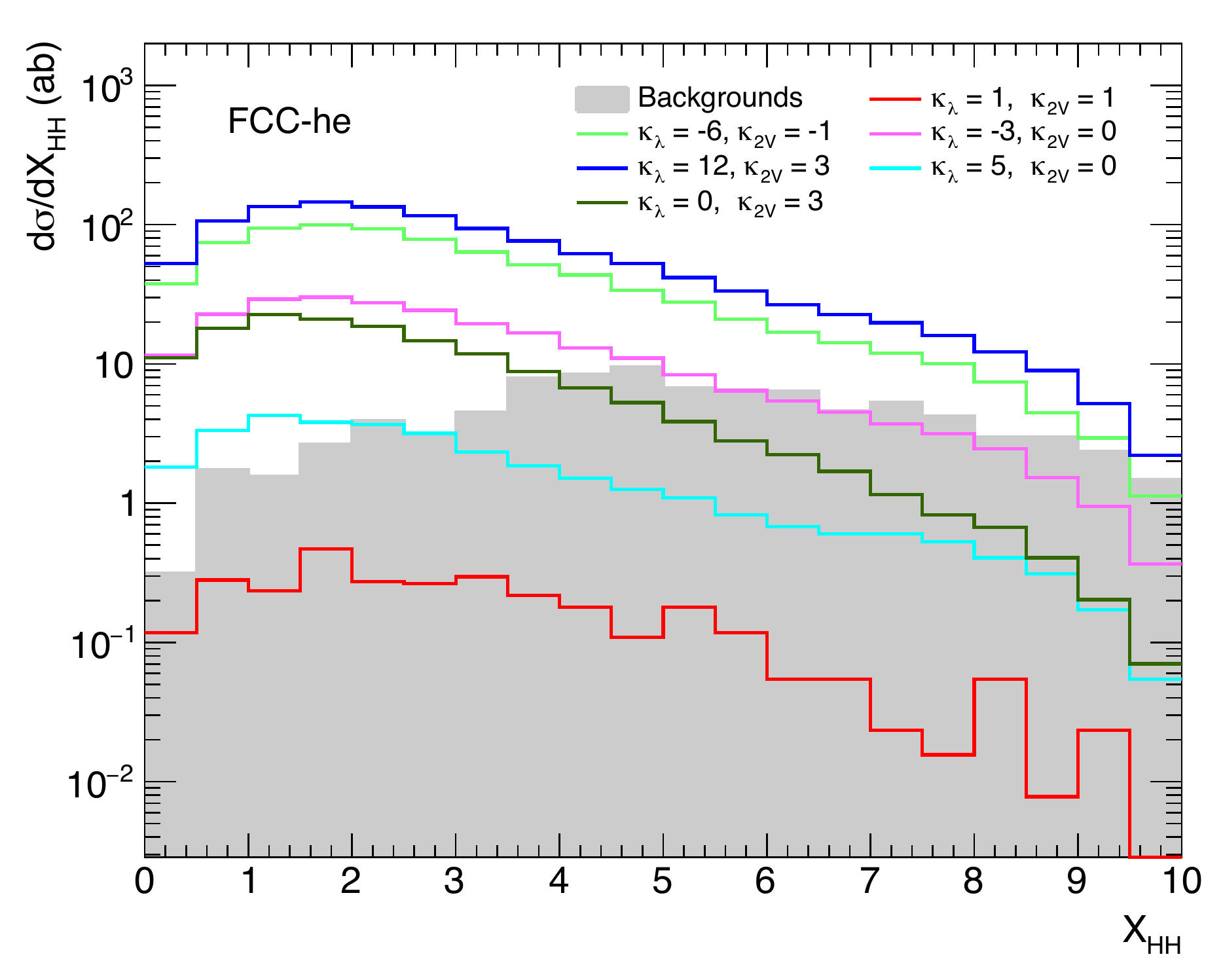}
    \caption{The differential cross-sections as a function of $X_{HH}$ at the LHeC (upper panel) and FCC-he
    (lower panel)
    with the unpolarized electron beam.  
    The total background is shown as the histogram in gray.
The six different signal hypotheses 
are  $(\kappa_{\lambda}, \kappa_{2V}) = \{(-6,-1), (12, 3), (0,3), (1,1), (-3,0), (5,0)\}$ shown in green, blue, olive, red, purple,     and cyan respectively.}
    \label{fig:XHH}
\end{figure}

In this section, we update the ATLAS analysis strategy 
for the VBF production of $HH$ \cite{Aad:2020kub},
to optimize the signal significance at the LHeC and FCC-he. 
As summarized in Table \ref{tab:Efficiency},
the event selections take the following steps:  
\bit
\item \textit{Initial:}\\
The initial number of events, $n_0$, is obtained from the full detector-level simulation.
We consider the decays of $H \to \bb$, $Z\to\bb$, and
both the semi-leptonic and hadronic decays of a top quark pair.
\item \textit{$4b$-tag:}\\
We require the presence of at least four 
$b$-tagged jets with $p_T^b > 20~$GeV and $|\eta^b| < 5$.
The acceptance times efficiency for the signal processes is around $10$-$16\%$,
depending on the values of $\kappa_\lambda$ and $\kappa_{2V}$.
%
\item \textit{Forward jet:}\\
We demand that at least one jet, untagged as a $b$ jet,
has $p_T^{j_{\rm f}} > 20~$GeV and  
$1.5 < \eta^{j_{\rm f}} < 7$.
Note that the definition of being forward at asymmetric $e^- p$ colliders
is different from that at the LHC. 
This selection reduces the signal events
by about $20\%$, irrespective of the hypothesis of $\kptv$ and $\kplm$.
\item \textit{Lepton veto:}\\
We veto the events which contains an isolated lepton ($\ell=e^\pm,\mu^\pm$)
with $p_T^\ell > 10~$GeV and $|\eta^{\ell}| < 5$. 
The criteria of \emph{lepton isolation} is required so that charged leptons from heavy hadron decays are not subject to this selection but their momenta are added to the hadronic jet if $\Delta R(\ell, {\rm jet}) < 0.2$.
Here $\Delta R\equiv \sqrt{\Dt\eta^2 + \Dt\phi^2}$.
This selection is very effective in suppressing the NC backgrounds.
The event yields for the signal and CC backgrounds remain almost the same.
\item \textit{$E_{T}^{\rm miss}$-cut:}\\
This selection consists of two requirements, $E_{T}^{\rm miss} > 40~$GeV and
$|\phi_{j_{\rm f}} - \phi_{E_{T}^{\rm miss}}| > 0.4$.
The latter removes the backgrounds with incorrectly measured $E_{T}^{\rm miss}$.
At this stage, the signal event yield is reduced by about $10\%$.
\item \textit{Minimum $D_{HH}$:}\\
The mission here is to find two Higgs boson candidates from four $b$-tagged jets.
There are three possible combinations for pairing two $b$-jets out of four, called the dijet.
In each combination,
we order two dijets according to their transverse momentum,
and call them the `leading` dijet and the `sub-leading` dijet.
Computing the angular separation of two $b$-jets inside each dijet system,
$\Delta R_{\rm lead}$ and $\Delta R_{\rm slead}$, 
we require
\begin{eqnarray}
\label{eq:DR}
\Delta R_{\rm lead} < && \!\!\!\!\!\!\!\!\!\!
\begin{cases}
		\frac{653\gev}{M_{4b}} + 0.475 & \mathrm{if}\ M_{4b} < 1250\,{\rm GeV}, \\
		1.0\, & \mathrm{if} \ M_{4b} > 1250\,{\rm GeV},
		\end{cases} \\ \nn
 \Delta R_{\rm slead} < && \!\!\!\!\!\!\!\!\!\!
 \begin{cases}
	\frac{875\gev}{M_{4b}} + 0.35 & ~~\mathrm{if}\ M_{4b} < 1250\,{\rm GeV}, \\
	1.0 & ~~\mathrm{if}\ M_{4b} > 1250\,{\rm GeV}, 
	\end{cases}
\end{eqnarray}
where $M_{4b}$ is the invariant mass of the four $b$-tagged jets.
Among the pairings that satisfy Eq.~(\ref{eq:DR}), we 
choose the pairing with the smallest value of $D_{HH}$ as the final $HH$ candidate.
Here $D_{HH}$ is~\cite{Aad:2020kub}
\begin{eqnarray}
\label{eq:DHH}
D_{HH} \!\!\!&=&\!\!\! \sqrt{(M_{\rm dijet}^{\rm lead})^2 + (M_{\rm dijet}^{\rm slead})^2} \\ \nn
\!\!\!&\times&\!\!\!
 \left|\sin\left(\tan^{-1}\frac{M_{\rm dijet}^{\rm slead}}{M_{\rm dijet}^{\rm lead}} - \tan^{-1}\frac{116.5~{\rm GeV}}{123.7~{\rm GeV}}\right)  \right|,
\end{eqnarray} 
where $M_{\rm dijet}^{\rm lead}~\lf M_{\rm dijet}^{\rm slead} \ri$ 
is the invariant mass of the leading (sub-leading) dijet system.
The values of $116.5~$GeV and $123.7~$GeV are adopted 
to properly treat the energy loss in the semi-leptonic decays of the $b$-hadrons.
\item \textit{$X_{HH}$-cut:}\\
Finally, the signal region is defined by the following variable~\cite{Aad:2020kub}:
\bea
\label{eq:XHH}
\!\!\!\!\!\!\!\!\!\!\!\!\!\!&& \!\!\!\!\!\! X_{HH} 
\\[3pt] \nn
\!\!\!\!\!\!\!\!\!\!\!\!\!\!&& \!\!\!\!\!\! \equiv \!\!
\sqrt{\left(\frac{M_{\rm dijet}^{\rm lead} - 123.7~{\rm GeV}}{11.6\gev}\right)^2 + \left(\frac{M_{\rm dijet}^{\rm slead} - 116.5~{\rm GeV}}{18.1\gev}\right)^2}.
\eea
The ATLAS collaboration required $X_{HH} <1.6$ to maximize the LHC signal significance. 
To optimize the search at the LHeC and FCC-he,
we present the differential cross-sections as a function of $X_{HH}$ 
for the LHeC and FCC-he in Fig.~\ref{fig:XHH}.  
The histograms in gray represent the total background distributions.
We also show the signal results in six different hypotheses of
    $(\kappa_{\lambda}, \kappa_{2V}) = \{(-6,-1), (12, 3), (0,3), (1,1), (-3,0), (5,0)\}$ in green, blue, olive, red, purple,
    and cyan respectively.
It is clear to see that the backgrounds are distributed in the high $X_{HH}$ region.
We have calculated the signal significance, to be defined below,
for different values of the upper-cut on $X_{HH}$.
We found that $X_{HH}<3$ ($X_{HH}<2$) at the LHeC (FCC-he) maximizes the signal significance,
by which we define the signal region.
Note that especially at the LHeC,
$X_{HH}<3$ allows significantly more data in the signal region than the LHC cut of $X_{HH}<1.6$,
which partially offsets the weakness of the LHeC's having tiny signal events.
\eit

\subsection{Results}

\begin{figure}[!h]
    \centering
    \includegraphics[width=0.9\linewidth]{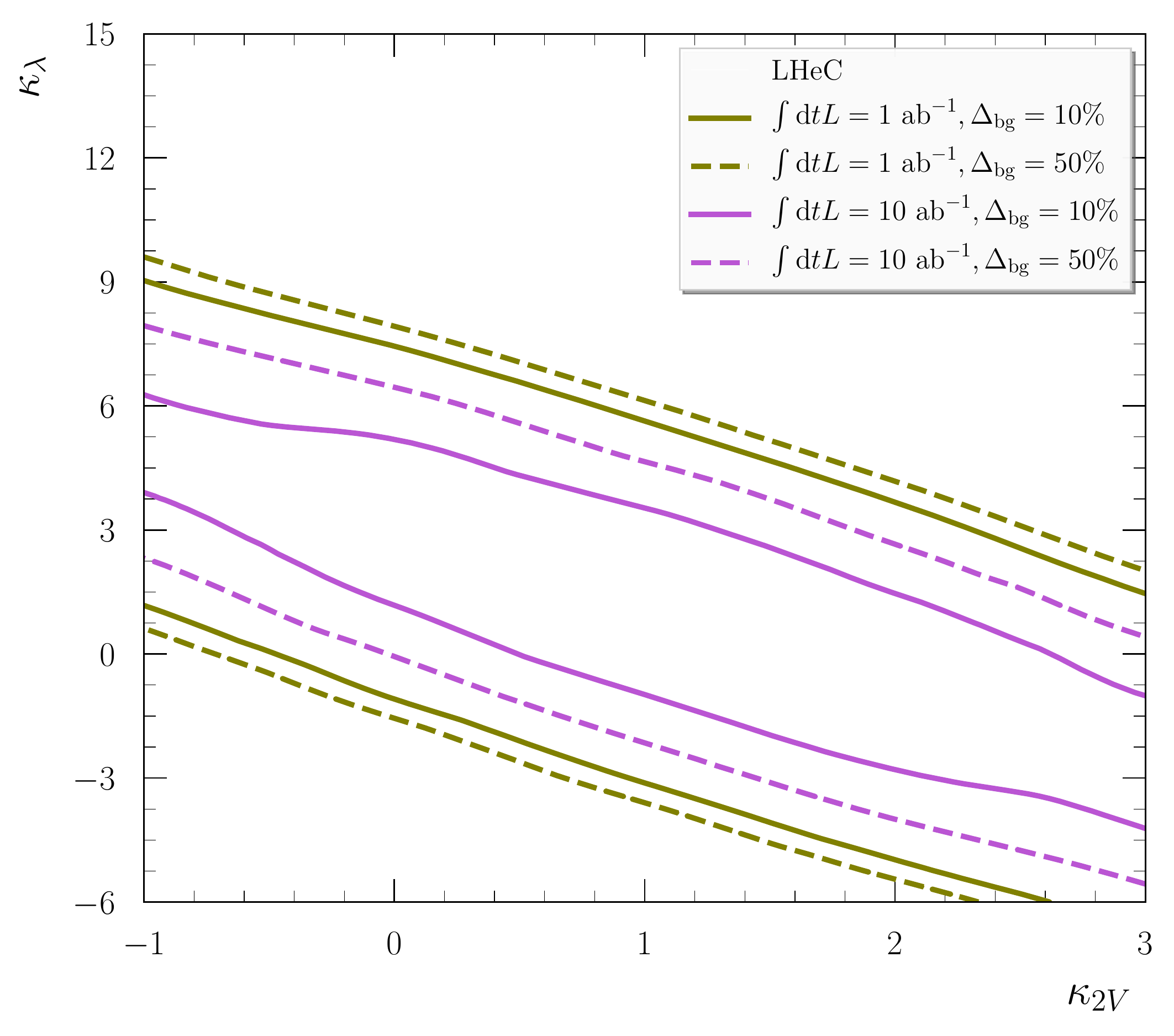}
    \hfill
    \includegraphics[width=0.9\linewidth]{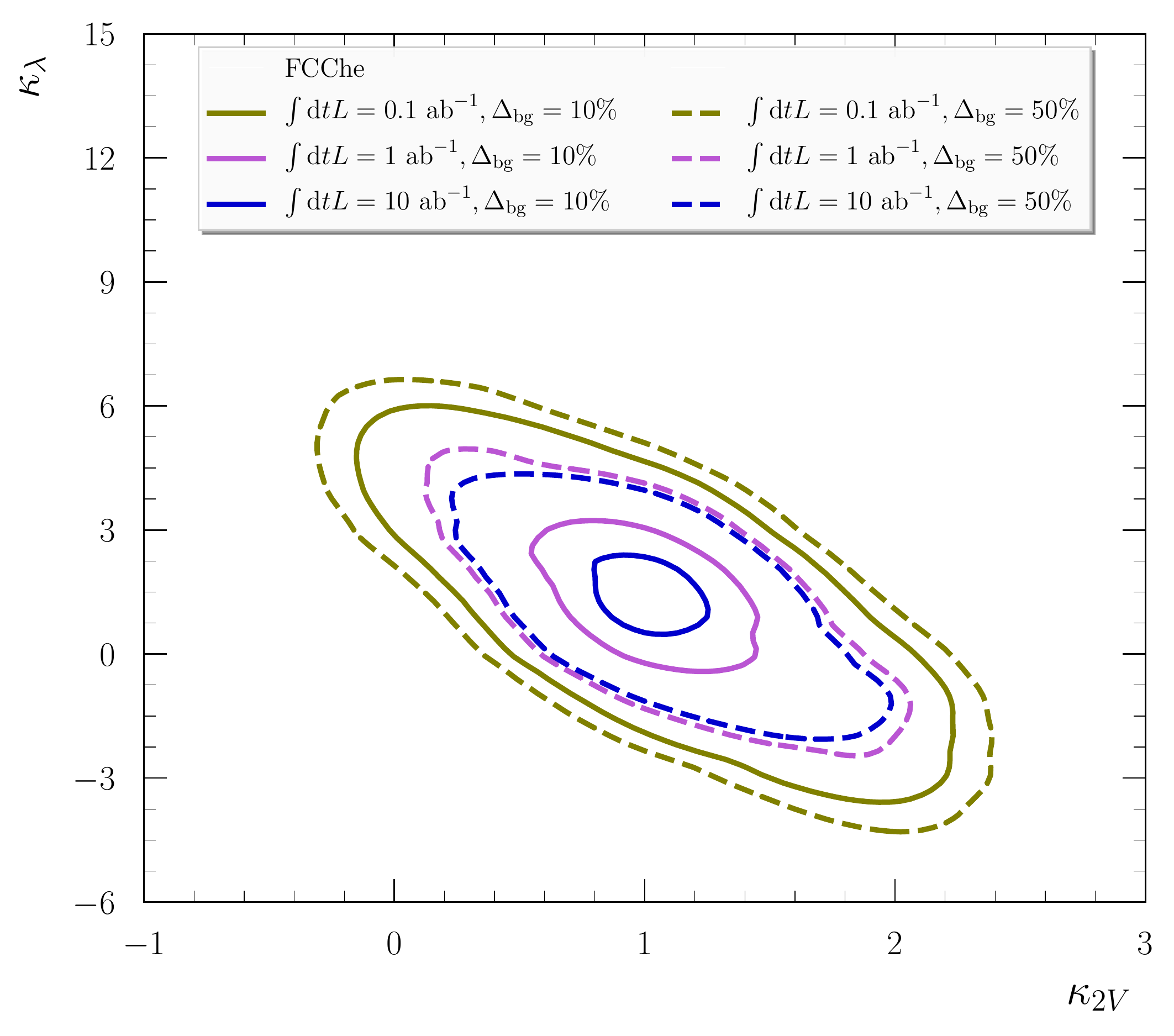}
    \caption{The expected exclusions from Higgs boson pair production projected on $(\kappa_{2V},\kplm)$
    at the LHeC and FCC-he with the electron beam polarization of $P_e=-80\%$. 
    We consider several options for the total integrated luminosity: 
    $1\iab$ (olive) and $10\iab$ (orchid) at the LHeC;
    $0.1~{\rm ab}^{-1}$ (olive), $1~{\rm ab}^{-1}$ (orchid), and $10~{\rm ab}^{-1}$ (blue) at the FCC-he.
Two choices of the background uncertainty are considered,
$\Dt_{\rm bg}=10\%$ (solid) and $\Dt_{\rm bg}=50\%$ (dashed).}
    \label{fig:significance}
\end{figure}

In this section, we discuss the results of our analysis. 
After the full selection, the signal efficiency
is about $2.0\%$ for the LHeC and about $1.4\%$ at the FCC-he, while the background
efficiency is about $\mco(10^{-4}$-$10^{-3})\%$ (see Table \ref{tab:Efficiency}).
To obtain the discovery potential, 
we compute the signal significance including the background uncertainty~\cite{Cowan:2010js},
defined by
\bea
\label{eq:significance}
\mathcal{S} \!\!\! &=& \!\!\!  
\Bigg[2(N_s + N_b) \log\left(\frac{(N_s + N_b)(N_b + \delta_b^2)}{N_b^2 + (N_s + N_b)\delta_b^2} \right) 
\\ \nn  && \quad - 
\frac{2 N_b^2}{\delta_b^2} \log\left(1 + \frac{\delta_b^2 N_s}{N_b (N_b + \delta_b^2)}\right)\Bigg]^{1/2},
\eea
where $N_s$ is the number of signal events, $N_b$ is the number of background events, 
and $\delta_{b} = \Delta_{\rm bg} N_b$ is the uncertainty in the background yields. 
The numbers of the signal and background events are 
\begin{eqnarray}
\!\!\! N_s \!\!\! &=& \!\!\! \mathcal{L}_\tot\times \epsilon_{HH}\sigma_{HH} \br_{H\to b\bar{b}}^2, 
\\ \nn
N_b \!\!\! &=& \!\!\! 
\mathcal{L}_\tot\times \bigg[\epsilon_{bbjj} \sigma_{bbjj} + \epsilon_{ZZ} \sigma_{ZZ} \br_{Z\to b\bar{b}}^2
+\epsilon_{Hb\bar{b}} \sigma_{Hb\bar{b}} \br_{H\to b\bar{b}}
\\ \nn \!\!\! && \!\!\!
 + \, 2 \epsilon_{HZ} \sigma_{HZ} \br_{H\to b\bar{b}} \br_{Z\to b\bar{b}} 
     + \epsilon_{Zb\bar{b}} \sigma_{Zb\bar{b}} \br_{Z\to b\bar{b}} 
  \\ \nn \!\!\! && \!\!\!
 + \, \epsilon_{t\bar{t}} \sigma_{t\bar{t}} \br_{t\to bjj}^2
 + 2 \epsilon_{t\bar{t}} \sigma_{t\bar{t}} \br_{t\to bjj} \br_{t\to b\ell\nu}\Bigg],
\end{eqnarray}
where $\mathcal{L}_\tot$ is the total integrated luminosity, 
$\epsilon_X$ is the acceptance times efficiency for the process $X$ in the signal region,
and $\br_X$ is the branching ratio of the decay $X$.  
Brief comments on the error estimation for the backgrounds are in order here.
In principle,
the background errors show different variation according to 
jet energy scale, the momentum smearing, 
$b$-tagging efficiency, jet energy resolution, and theoretical uncertainties.
Since the detailed study is beyond the scope of this work,
we take two simple cases, $\Delta_{\rm bg} = 10\%$ and $\Delta_{\rm bg} = 50\%$.\footnote{In this study,
we adopted a conservative approach for the background uncertainties.
Considering the expected improvement of various precisions,
e.g., the PDF precision at the LHeC,
we expect that $\Delta_{\rm bg} = 10\%$ can be obtained in the future.}

In Fig. \ref{fig:significance}, we display the expected exclusions on the plane of $\kappa_{2V}$ and 
$\kappa_{\lambda}$ at the LHeC (upper panel) and the FCC-he (lower panel),
corresponding to $\mathcal{S} > 2$.  
We consider the electron beam polarization of $P_e=-80\%$ and two cases of the background uncertainty, 
$\Dt_{\rm bg}=10\%$ (solid) and $\Dt_{\rm bg}=50\%$ (dashed).
For the total integrated luminosity $\mathcal{L}_\tot$,
we take $1\iab$ (olive) and $10\iab$ (orchid) at the LHeC,
and $0.1~{\rm ab}^{-1}$ (olive), $1~{\rm ab}^{-1}$ (orchid), and $10~{\rm ab}^{-1}$ (blue) at the FCC-he.
The common result of the LHeC and FCC-he
is that the same-sign $\kptv$ and $\kplm$ region
is more strongly constrained because of the constructive interference discussed before.

In detail, the LHeC and FCC-he have different exclusion potential.
In general, the LHeC has limitations in constraining $\kappa_{2V}$ and $\kplm$ 
because of its lower center-of-mass energy.
Nevertheless, it can produce some meaningful results.
If $\kptv=1$,
the LHeC data with $\mathcal{L}_\tot = 1~{\rm ab}^{-1}$ and $\Delta_{\rm bg}=10\%$
can constrain $\kplm$ as $-3 \lsim \kplm \lsim 6$,
which is weaker than the HL-LHC prospect.
If $\kplm=1$, 
the LHeC data with ${\cal L}_{\rm tot} = 1~{\rm ab}^{-1}$ and $\Delta_{\rm bg} = 10\%$
can exclude $\kptv\lsim -1$ and $\kptv \gsim 3.2$,
which is compatible with the current bound on $\kptv \in [-0.66,2.89]$
at 95\% C.L.~\cite{Aad:2020kub}.
Considering the feasibility of the concurrent operation of the HL-LHC and LHeC,
two colliders shall play a complementary role in probing $\kptv$.
In terms of the ratio of the cross-section of the CC VBF production of $HH$ to the SM value,
the LHeC with $\mathcal{L}_\tot=1\iab$ and $\Dt_{\rm bg}=10~(50)\%$
can limit ${\sigma}/{\sigma_{\rm SM}}\lsim 30~(35)$.

On the other hand, 
the FCC-he has high potential in probing both $\kptv$ and $\kplm$.
For $\kplm \in [0.1,2.3]$ suggested by the HL-LHC prospect study~\cite{Cepeda:2019klc},
$|\kptv| \gsim 0.2$ is to be excluded by the FCC-he data 
with $\mathcal{L}_\tot = 10~{\rm ab}^{-1}$ and $\Delta_{\rm bg} = 10\%$.
Two important reasons for this high precision are higher signal cross-section
and similar rejection rates of the SM backgrounds (see Table \ref{tab:Efficiency}).
At the FCC-he with $\Delta_{\rm bg}=10\%~(50\%)$,
we estimated conservative bounds on the ratio of the Higgs pair production cross section 
to the SM value as follows:
\begin{eqnarray}
\left. \frac{\sigma}{\sigma_{\rm SM}} \right|_{\rm FCC-he}< 
\left\{
{\renewcommand{\arraystretch}{1.2} 
 \begin{array}{ll}
		11~(14) & \mathrm{for}~ \mathcal{L}_\tot = 0.1~{\rm ab}^{-1}, \\
		3.5~(8) & \mathrm{for}~ \mathcal{L}_\tot = 1~{\rm ab}^{-1},\\
		1~(7) & \mathrm{for}~ \mathcal{L}_\tot = 10~{\rm ab}^{-1}. \\
		\end{array}
		}
		\right.
\end{eqnarray}

Final comments on the role of higher electron beam energy in probing
the $HH$ process are in order here. Although it is practical
for the LHeC working group to choose $E_{e}=50{\;{\mathrm{GeV}}}$ due to the
cost issues, the physics gain from higher $E_{e}$ is more important than
anything else. We found that setting $E_{e}=120{\;{\mathrm{GeV}}}$ increases
the background cross sections by a factor of $2.32$~($1.82$) at the LHeC~(FCC-he).
For the signal cross sections, the enhancement factor is $2.1$ --
$2.7$ at the LHeC and $4.4$ -- $5.9$ at the FCC-he, depending on the values
of $\kappa _{\lambda }$ and $\kappa _{2 V}$. Assuming similar efficiencies
for both the signal and backgrounds to those in Table~\ref{tab:Efficiency}, we expect that
the significance increases by a factor of $1.4$ -- $1.8$ ($3.3$ --
$4.4$) at the LHeC (FCC-he). At the FCC-he, increasing $E_e$ into
$120{\;{\mathrm{GeV}}}$ has almost the same effect as increasing the total
luminosity tenfold. We strongly suggest that the FCC-he working group seriously
consider the higher $E_{e}$ option.

\section{Conclusions}
\label{sec:Conclusions}

Upon the current status where both the trilinear Higgs self-coupling modifier
($\kappa _{\lambda }$) and the quartic coupling modifier between a Higgs
boson pair and a vector boson pair ($\kappa _{2V}$) are unmeasured, we
consider two electron-proton colliders, the LHeC and FCC-he, in probing
$\kappa _{\lambda }$ and $\kappa _{2V}$ simultaneously. As a proton-proton
collider, the LHC cannot avoid the gluon fusion pollution in the VBF production of a Higgs
pair, which becomes much worse for
$\kappa _{\lambda }\neq 1$. At electron-proton colliders, the gluon fusion
pollution is absent, and thus the charged-current VBF production of a Higgs
boson pair can be solely measured if there is enough signal significance.
With this motivation, we study the detailed phenomenology of
$p e^{-} \to HHj\nu _{e}$ in the ${b \bar{b}}{b \bar{b}}$ final state
and suggest a search strategy at the LHeC and FCC-he based on the full
simulation. Taking the default CDR values, we took
$E_{e}=50~(60){\;{\mathrm{GeV}}}$ and $E_{p}=7~(50){\;{\mathrm{TeV}}}$ at the LHeC
(FCC-he).

First, we calculated the parton-level cross-sections of the signal in the
parameter space of $(\kappa _{2V},\kappa _{\lambda })$ as well as all relevant
backgrounds. Theoretical uncertainties from the variations of the scales
and PDF are also calculated. Although the backgrounds are relatively manageable,
the SM cross-section ($\kappa _{2V}=\kappa _{\lambda }=1$) is extremely small:
without including the Higgs boson decays,
$\sigma _{\mathrm{SM}} = 5.97{\;{\mathrm{ab}}}$ at the LHeC and
$\sigma _{\mathrm{SM}} = 233.77\;{\mathrm{ab}}$ at the FCC-he for the unpolarized
electron beam. It is very challenging to measure this process for the SM
values of $\kappa _{\lambda }$ and $\kappa _{2V}$. What is hopeful is that
a small deviation from $\kappa _{\lambda }=\kappa _{2V}=1$ greatly enhances
the signal rate. The electron-proton collider can exclude a large portion
of the $(\kappa _{2V},\kappa _{\lambda })$ space.

We have completed the analysis with full simulations to devise an optimal
strategy. We found that most of the current ATLAS search strategies for
$HH$ via VBF production apply to those at the LHeC and FCC-he. The key
difference of ours is the cut on $X_{HH}$, defined in Eq.~(\ref{eq:XHH}).
We found that the signal significance at the LHeC (FCC-he) is maximized
by $X_{HH}<3$ ($X_{HH}<2$). A larger upper bound on $X_{HH}$ than the one
for the LHC, $X_{HH}^{\rm LHC}<1.6$, increases the signal significance. As the final result, we calculated the expected
exclusions on
$(\kappa _{2V},\kappa _{\lambda })$. The LHeC can play a meaningful role
in probing $\kappa _{2V}$: the data with the total integrated luminosity
of $\mathcal{L}_{\mathrm{tot}}= 1~{\mathrm{ab}}^{-1}$ and the background uncertainty
of $\Delta _{\mathrm{bg}}=10\%$ can constrain
$-1 \lesssim \kappa _{2V}\lesssim 3.2$ for $\kappa _{\lambda }=1$. The FCC-he
has immense power in constraining both $\kappa _{2V}$ and
$\kappa _{\lambda }$. If $\kappa _{\lambda }\in [0.1,2.3]$ as the HL-LHC prospect,
$|\kappa _{2V}| \gtrsim 0.2$ is to be excluded with
$\mathcal{L}_{\mathrm{tot}}= 10~{\mathrm{ab}}^{-1}$ and
$\Delta _{\mathrm{bg}} = 10\%$. We hope that this study would provide input
to strongly support the future programs of electron-proton colliders which
are capable of measuring two fundamental couplings,
$\kappa _{2V}$ and $\kappa _{\lambda }$.

\section*{Acknowledgments}
The authors would like to thank Mukesh Kumar and Xifeng Ruan for stimulating discussions and for providing 
the necessary material to reproduce their results. AJ would like to thank Oliver Fischer for pointing out to the
updated \texttt{Delphes} cards for the LHeC and FCC-he. This work is supported by the National Research Foundation 
of Korea, Grant No. NRF-2019R1A2C1009419. 


\end{document}